\documentclass[letterpaper]{article} 
\usepackage{aaai23}  
\usepackage{times}  
\usepackage{helvet}  
\usepackage{courier}  
\usepackage[hyphens]{url}  
\usepackage{graphicx} 
\usepackage{natbib}  
\usepackage{caption} 
\usepackage{algorithm}
\usepackage{algorithmic}

\usepackage{newfloat}
\usepackage{listings}
\DeclareCaptionStyle{ruled}{labelfont=normalfont,labelsep=colon,strut=off} 
\floatstyle{ruled}
\newfloat{listing}{tb}{lst}{}
\floatname{listing}{Listing}
\title{DiPPS: Differentially Private Propensity Scores for Bias Correction}
\author {
    Liangwei Chen\equalcontrib\thanks{Work done while at EPFL.}\textsuperscript{\rm 1},
    Valentin Hartmann\equalcontrib\textsuperscript{\rm 2},
    Robert West\textsuperscript{\rm 2}
}
\affiliations {
    \textsuperscript{\rm 1} Google\\
    \textsuperscript{\rm 2} EPFL\\
    chenlw@google.com, valentin.hartmann@epfl.ch, robert.west@epfl.ch
}

\usepackage[utf8]{inputenc} 
\usepackage{booktabs}       
\usepackage{amsfonts}       
\usepackage{nicefrac}       
\usepackage{microtype}      

\usepackage{subcaption}
\usepackage{mathptmx}

\usepackage{hyphenat}
\usepackage{xspace}

\newcommand{\xhdr}[1]{\vspace{1.7mm}\noindent{{\bf #1.}}}

\newcommand{\eqnref}[1]{Eq.~\ref{#1}}

\newcommand{\Tabref}[1]{Table~\ref{#1}}

\newcommand{\Figref}[1]{Fig.~\ref{#1}}

\usepackage{amsthm}
\theoremstyle{plain}
\newcounter{thmcounter}
\newcounter{assumptioncounter}

\newtheorem{definition}[thmcounter]{Definition}

\newtheorem{assumption}[assumptioncounter]{Assumption}

\usepackage{mathtools}

\usepackage{amsmath}

\usepackage[normalem]{ulem} 

\newcommand*{\msetl}{\{\mskip-4mu\{}
\newcommand*{\msetr}{\}\mskip-4mu\}}

\usepackage{cryptocode}

\newcommand\restr[2]{{
  \left.\kern-\nulldelimiterspace 
  #1 
  \right|_{#2} 
  }}
  
\usepackage[rightcaption]{sidecap}

\begin{document}

\maketitle

\begin{abstract}
In surveys, it is typically up to the individuals to decide if they want to participate or not, which leads to participation bias: the individuals willing to share their data might not be representative of the entire population. Similarly, there are cases where one does not have direct access to any data of the target population and has to resort to publicly available proxy data sampled from a different distribution. In this paper, we present Differentially Private Propensity Scores for Bias Correction (\textit{DiPPS}), a method for approximating the true data distribution of interest in both of the above settings. We assume that the data analyst has access to a dataset $\widetilde{\mathcal{D}}$ that was sampled from the distribution of interest in a biased way. As individuals may be more willing to share their data when given a privacy guarantee, we further assume that the analyst is allowed locally differentially private access to a set of samples $\mathcal{D}$ from the true, unbiased distribution. Each data point from the private, unbiased dataset $\mathcal{D}$ is mapped to a probability distribution over clusters (learned from the biased dataset $\widetilde{\mathcal{D}}$), from which a single cluster is sampled via the exponential mechanism and shared with the data analyst. This way, the analyst gathers a distribution over clusters, which they use to compute propensity scores for the points in the biased $\widetilde{\mathcal{D}}$, which are in turn used to reweight the points in $\widetilde{\mathcal{D}}$ to approximate the true data distribution. It is now possible to compute any function on the resulting reweighted dataset without further access to the private $\mathcal{D}$. In experiments on datasets from various domains, we show that \textit{DiPPS} successfully brings the distribution of the available dataset closer to the distribution of interest in terms of Wasserstein distance. We further show that this results in improved estimates for different statistics, in many cases even outperforming differential privacy mechanisms that are specifically designed for these statistics.
\end{abstract}

\section{Introduction}
\label{sec:intro}
Participation bias is the bias that occurs when the non\hyp participation of certain individuals in a study biases the collected data. Participation bias has been identified as a problem in surveys in many different domains:
\begin{itemize}
    \item In a sexuality survey \cite{dunne1997participation}, participants had higher levels of education, were less politically conservative, less harm\hyp avoidant and more sexually liberal than non-participants.
    \item In a longitudinal health study \cite{lissner2003participation}, individuals who not only participated in the initial, but also in later stages of the study were better educated and more healthy than individuals who dropped out during the study period.
    \item In a study on sick leave of employees \cite{van1999nonresponse}, people who were on sick leave less often and shorter were more likely to participate than others.
\end{itemize}

Participation bias may also occur outside of traditional surveys. Nowadays many software products ask their users whether they want to share anonymous usage statistics. Take the example of a browser developer that wants to collect data about how often different features of their browser are used. Users who refuse to share their data with the browser developer (non\hyp participants) might do this because they are more concerned about their privacy than users who do share their data (participants) \cite{korkeila2001non,ronmark1999non}. Non-participants would hence be more likely to use privacy\hyp related features such as tracking protection or the private browsing mode. Consequently, the browser developer would underestimate the use of such features if they would base their analysis solely on the data of the participants.
Note that the promise of collecting \emph{anonymous} usage statistics is not of much worth to users, since their identity is typically still known to the browser developer, e.g., via the account with which they logged into the browser, via their IP address, etc. Even if the developer does not associate the collected records with the identity of the user, users are still at risk of de\hyp anonymization \cite{narayanan2008robust,sweeney2000simple}.

A related problem occurs when initially no data from the target population exists at all and one has to resort to proxy data. E.g., a linguist might want to study language patterns in private messages, but only has access to public Twitter message. The developer of a camera app for smartphones might want to improve the post\hyp processing algorithms by analyzing the most typical lighting conditions in their users' photos, but since the photos are stored locally, the developer must resort to publicly available photos on platforms like Flickr. In these cases, the public proxy data and the target data come from the same domain, but differ in their distribution: Private messages contain more intimate information than public ones and people select only their most beautiful photos to upload to Flickr.

For this reason, big efforts are undertaken to convince more individuals to participate in studies \cite{de2005evaluation}. A particularly convincing argument for participation might be the promise of local differential privacy. Users might be more willing to share, e.g., a scalar differentially private value than a full plain\hyp text vector with information about themselves \cite{warner1965randomized}.
A meta-analysis by \citet{lensvelt2005meta} based on \(38\) studies shows that individuals are more prone to providing correct answers to survey questions when given a differential privacy guarantee via the randomized response mechanism \cite{warner1965randomized}.
Companies such as Google \cite{erlingsson2014rappor}, Apple \shortcite{apple2017learning} and Microsoft \cite{ding2017collecting} have recognized this potential and implemented data collection mechanisms in their products that provide local differential privacy.

In this paper, we assume that there are two sets of individuals: participants, to whose data we have full access, and non-participants, to whose data we only have locally differentially private access (for individuals to whose data we do not even have locally differentially private access, see `Problem Definition'). Our method, Differentially Private Propensity Scores for Bias Correction (\textit{DiPPS}), uses the differentially private access to the non-participants' data to estimate the data distribution of all individuals. It reduces the participation bias that would occur from using only the participants' data for drawing conclusions about the entire population. Our method can even be used when there exist no participants; then, the participant data is replaced by a proxy dataset, and only the non\hyp participants' data distribution is approximated. The differentially private value that the non\hyp participants share with the data analyst is the value of a single categorical variable and requires only one round of communication. This makes \textit{DiPPS} suitable even for offline settings such as offline surveys, and further makes it easy to explain what data is shared and how privacy is preserved to laymen.

\xhdr{Overview of \textit{DiPPS}}
Our method consists of three main steps.
\begin{enumerate}
    \item First, a clustering model is trained on the participant data. This model transforms each data point into a probability distribution over a finite number of classes. In our case, this is a probabilistic clustering model, but other implementations such as dimensionality reduction models are possible as well.
    \item Then, this model is shipped to the non\hyp participants. They apply the clustering model to their data and sample a single value from the resulting probability distribution in a locally differentially private way. Afterwards, they return this value. 
    \item In the last step, the values returned by the non\hyp participants are used to estimate the propensity of each of the participants' data points to be indeed part of the participant dataset. These propensity scores are then used to reweight the participant data points to either model the distribution of all individuals, or the distribution of only the non\hyp participants in the case of a proxy dataset.
\end{enumerate}
The non\hyp participants are guaranteed local differential privacy:

\xhdr{Local differential privacy}
Differential privacy \cite{dwork2006calibrating} is a privacy notion that is widely used in academic research and increasingly also in industry applications. It states that the output of a randomized mechanism that is invoked on a database should reveal only very little information about any individual record in the database. Local differential privacy applies the concept to distributed databases, where each individual holds their own data \cite{kasiviswanathan2011can}:
\begin{definition}
Let \(\mathcal{M}\) be a randomized mechanism and let \(\varepsilon>0\). \(\mathcal{M}\) provides \(\varepsilon\)\hyp local differential privacy if, for all pairs of possible values \(x,x'\) of an individual's data and all possible sets of outputs \(S\):
\begin{equation*}
    \Pr[\mathcal{M}(x)\in S]\leq e^{\varepsilon} \Pr[\mathcal{M}(x')\in S].
\end{equation*}
\end{definition}

An \(\varepsilon\)\hyp differential privacy guarantee for a mechanism \(\mathcal{M}\) is an upper bound on the amount by which an adversary can update any prior belief about the database, given the output of \(\mathcal{M}\) \cite{kasiviswanathan2014semantics}. Smaller values of \(\varepsilon\) mean more privacy, larger values less privacy. A typical choice is \(\varepsilon=1\).

\xhdr{Overview of the paper}
We first discuss related work. Then we formally define the participation bias correction problem that \textit{DiPPS} solves, followed by the description of the different components of our solution (see \Figref{fig:diagram} for an overview) and the results of the various experiments that we use to evaluate it. Finally, we discuss limitations and possible extensions of \textit{DiPPS} and summarize the paper.

\section{Related Work}
\label{sec:related}
Traditional methods for participation bias correction \cite{lundstrom1999calibration,valliant1993poststratification,ekholm1991weighting} assume that one has auxiliary information about the respondents and the non-respondents. This could, e.g., be geographical information when doing a survey via house visits, or known population totals. If, for example, the target population is the entire population of a country, then the population totals can come from census data. If the covariates \(D\), the auxiliary information \(A\) and the variable \(Z\) indicating participation/non\hyp participation form the Markov chain \(D \text{--} A \text{--} Z\), then these methods can work well. See \cite{groves2006nonresponse} for more details.
Often methods for participation bias correction use propensity scores, which are the probabilities of the individuals being respondents, given their covariates \cite{little2003weighting}. The collected samples are then reweighted with the inverse of the propensity scores to correct for the participation bias. This is what we do in our method as well.

Reweighting data records can also be required for causal inference. \citet{agarwal2021causal} propose a reweighting method for estimating causal parameters such as the average treatment effect from data that has been released with differential privacy. Instead of weighting points with their inverse propensity score, they use an error-in-variable balancing technique. Like us in our concrete choice of implementation, they assume a low rank data matrix, and confirm the validity of this assumption on US census data.

Another related setting is the following: For a machine learning (ML) task, there exist two datasets, one is labeled and the other one unlabeled, and there is a covariate shift between the two. When training an ML model, one would like to account for this shift by also taking into account the unlabeled data. Several methods for solving this problem have been proposed \cite{huang2007correcting,rosset2005method,zadrozny2004learning}. In this context, the idea of using clustering to correct for sample selection bias has already been explored \cite{cortes2008sample}. All of these methods work only in the non-private setting, where one has direct access to the unlabeled data.

We, however, assume that we neither have auxiliary information about the non\hyp participants, nor non\hyp private access to parts of their data. Access to data of non\hyp participants is only allowed in a locally differentially private way. For providing local differential privacy in the processing of distributed data, there exists a multitude of methods: for computing means
\cite{wang2019collecting,duchi2018minimax}, for computing counts \cite{erlingsson2014rappor} or even for training machine learning models \cite{truex2019hybrid}, to just name a few. What all of these methods have in common is that each one only serves a single purpose. The data analyst has to decide beforehand which function they want to compute on the data. If they decide to perform additional analyses later, which is, e.g., the case in exploratory and adaptive data analysis, they have to invoke another differentially private mechanism. This requires further rounds of communication with the individuals that hold the data and, more importantly, each additional function computation decreases the level of privacy \cite{rogers2016privacy,kairouz2015composition}. As opposed to that, our method estimates a distribution. This means that once the method has been executed, the data analyst can compute any number of arbitrary functions they want on this distribution, including all kinds of statistics, but also more complicated functions such as training ML models. Furthermore, while methods for, e.g., locally differentially private gradient descent require many rounds of communication, our method works with a single round of communication.

A trust setting similar to ours has been introduced earlier by \citet{avent2017blender}. As opposed to us, they assume one dataset with locally differentially private access and one with centrally differentially private access, whereas we assume one dataset with locally differentially private access and one with non\hyp private access. They describe a method for computing the most popular records of a web search log \cite{avent2017blender} and methods for mean estimation \cite{avent2020power}, whereas we consider the more general problem of distribution estimation. Note that our method can in principle be extended to their setting with purely differentially privacy access to data; see our discussion section.

Other works \cite{kancharla2021robust,clark1998honest} consider bias in locally differentially private surveys due to users not following the DP protocol faithfully. This might occur in our setting if the data collecting party gives the users only the options to share their data without a privacy guarantee or with DP, but not to not share data at all. The authors propose to split the users into two groups, let those groups invoke DP mechanisms with different parameters, and compare the two sets of responses to correct for this bias.

\section{Problem Definition}
\label{sec:problem}
Our method is also applicable in an offline setting, but assume for simplicity that there is a company that is selling a software and wants to collect data from its users over the Internet to, e.g., analyze usage patterns to improve the software, train ML models that are to be integrated in the software, to spot market opportunities for new products, etc. We consider two settings:
\begin{enumerate}
    \item Users of the software get the option to share their data, e.g., usage statistics, as it is common in a lot of nowadays’ software (Windows, Firefox, ...). Some users decide to share their data directly (without a privacy mechanism in place), some decide to only share data with a local differential privacy guarantee. The company therefore has direct access to a (potentially biased) subset of the data and in addition it has locally differentially private access to the rest of the data.\label{problem:1}
    \item The company does not have direct access to any user data, but only to a proxy dataset that comes from a similar distribution as the user data. If the user data consists of private text messages, the proxy dataset could for example be tweets from Twitter or public forum posts. Assume that the company has locally differentially private access to the user data.\label{problem:2}
\end{enumerate}
In both settings, the company wants to use the data to which it has direct access to perform data analysis, ML model training or other data\hyp dependent tasks. But in both cases, that data is most likely biased: in \ref{problem:1}, the covariate distribution of users who are willing to share their data might differ from the covariate distribution of users who are not willing to share their data. In \ref{problem:2}, the data even comes from a different source. The problem that we are solving is the reduction of this bias. We now formalize this problem.

Let \(D\) be the random variable that subsumes the covariates of the user data.
Let \(Z\) be a binary random variable indicating whether the company has direct, non\hyp private accces to a data point or not.
This gives rise to the joint distribution \((D,Z)\). Assume that there exists a multiset \(X\) of samples \((d,z)\) of \((D,Z)\). Using \(\msetl\cdot\msetr\) to denote multisets, let \(X^0 = \msetl d\mid (d,0)\in X\msetr\) be the data to which the company only has locally differentially private access and let \(X^1 = \msetl d\mid (d,1)\in X\msetr\) be the data to which the company has direct access.
The goal is to estimate the distribution of \(D\) (in Setting \ref{problem:1}) or the distribution of \(D\mid Z=0\) (in Setting \ref{problem:2}).

In the following we will refer to the data that can be directly accessed (i.e., directly shared data in Setting \ref{problem:1} and proxy data in Setting \ref{problem:2}) as the participant data \(U_1\) and the data that can only be accessed in a locally differentially private way as the non\hyp participant data \(U_2\).
Note that we do not consider a third group of users: those who are not willing to share any data, not even when provided a privacy guarantee. We denote their data by \(U_3\). This third group of users is empty if the company previously collected the data of all users without any privacy mechanism in place and now offers the option for users to instead share only locally differentially private data, but not the option to share no data at all. In cases where the company gives users the option to share no data at all, this third group of users will not be empty and ignoring it will in many cases lead to bias. This is a general problem when giving users complete freedom of choice over sharing their data and not specific to our method. However, methods that can estimate \(D\) or \(D\mid Z=0\) are still is useful even in this case, because the difference between the distribution of \(U_1\cup U_2\) and the data of all users \(U_1\cup U_2\cup U_3\) will most likely be smaller than the difference between just \(U_1\) and \(U_1\cup U_2\cup U_3\).

\section{Proposed Solution}
\label{sec:solution}
\subsection{General Reweighting Framework}
\label{sec:reweighting}

Assume for now that we have access to (an approximation of) the propensity score function \(e(d) = \Pr(Z=1\mid D=d)\). We describe a method for approximating \(e(d)\) in the next subsection.

With the knowledge of \(X^1\) and the size of \(X^0\), we can approximate
\begin{equation*}
    \Pr(D=d\mid Z=1)\approx \frac{|\msetl d: d\in X^1\msetr|}{|X^1|}
\end{equation*}
and
\begin{equation*}
    \Pr(Z=0)\approx \frac{|X^0|}{|X^0|+|X^1|},\quad \Pr(Z=1)\approx \frac{|X^1|}{|X^0|+|X^1|}.
\end{equation*}
With these probabilities, we can compute
\begin{equation*}
    \Pr(D=d, Z=1) = \Pr(D=d\mid Z=1)\Pr(Z=1).
\end{equation*}
Hence,
\begin{align*}
    \Pr(Z=1\mid D=d) &= \frac{Pr(D=d, Z=1)}{\Pr(D=d)}\\
    &= \frac{\Pr(D=d\mid Z=1)\Pr(Z=1)}{\Pr(D=d)},
\end{align*}
and thus
\begin{equation*}
    \Pr(D=d) = \frac{\Pr(D=d\mid Z=1)\Pr(Z=1)}{\Pr(Z=1\mid D=d)}.
\end{equation*}

If instead of estimating \(\Pr(D=d)\) we want to estimate \(\Pr(D=d\mid Z=0)\), we can do this as follows:
\begin{align*}
    &\Pr(D=d\mid Z=0) = \frac{\Pr(D=d,Z=0)}{\Pr(Z=0)}\\
    &= \frac{\Pr(D=d) - \Pr(D=d,Z=1)}{\Pr(Z=0)}\\
    &= \frac{\Pr(D=d) - \Pr(D=d\mid Z=1)\Pr(Z=1)}{\Pr(Z=0)}.
\end{align*}

Note that we only need the propensity scores for the points in \(X^1\), because these are the only points that we have direct access to and thus the only points that we reweight.

\subsection{Propensity Score Computation}
\label{sec:prop_scores}

To apply the reweighting described in the previous subsection, we need to know the propensity scores \(e(d) = \Pr(Z=1\mid D=d)\). For computing them, we rely on the following assumption:
\begin{assumption}
\label{assumption}
\(D\) can be well approximated by a mixture model with \(K\) components or classes (e.g., clusters or LDA topics), and the participant and non\hyp participant data differ only in the mixture weights.
\end{assumption}
We denote the class membership random variable with values in \(\{1,\dots,K\}\) by \(C\). Each point \(d\) has an associated distribution \(C\mid D=d\) over the classes, approximated by a probability vector \(\rho^d\). The idea is to compute the distributions \(C\mid Z=0\) and \(C\mid Z=1\) of classes in \(X^0\) and \(X^1\), respectively, and use these distributions to compute the propensity scores for the points in each class. \textit{DiPPS} approximates the propensity scores in a locally differentially private way via a multi\hyp step procedure:
\begin{enumerate}
    \item Train a model on \(X^1\) that learns the \(K\) different classes and assigns to each point \(d\) that it is invoked on a probability distribution \(\rho^d\) over the classes. This model might for example be an LDA topic model, and \(\rho^d\) could be the topic vector for document \(d\) that is normalized to a probability distribution.
    \item Send the model to the non\hyp participating users, i.e., to those with data \(X^0\).
    \item Each of the non\hyp participating users uses the model to compute \(\rho^d\) for their data point \(d\). They then use \(\rho^d\) to sample and return one of the \(k\) classes, by invoking the exponential mechanism with \(\rho^d\) as a utility function (we explain the exponential mechanism below).\label{enum:exponential}
    \item Collect the noisy class counts and apply the postprocessing as described later in this subsection to estimate the distribution of \(C\mid Z=0\), i.e., which fraction of the non\hyp participants lies in each class.
    \item Compute a propensity score for each class (how likely is it that \(Z=1\) for a point in a given class) and use these class propensity scores to compute propensity scores for all \(d\in X^1\) based on \(\rho^d\).
\end{enumerate}

We now describe the different steps in detail. We will start with the model that learns the \(K\) classes, continue with the exponential mechanism and the postprocessing of its outputs, and end with the computation of the propensity scores.

\xhdr{Implementation of the class assignment}
For our experiments we train a clustering model on the participant data to compute class membership probabilities, where the classes in our case are clusters. The model consists of a dimensionality reduction via PCA, followed by a Gaussian mixture model (GMM) with \(K\) components. This model could be replaced by any other model that can learn different classes in a dataset, such as other clustering algorithms.
The debiasing performance of our method depends to a large part on how well the clustering algorithm can learn the different clusters in the data. Fortunately, the problem of clustering has been studied for decades, and there exists a wide variety of algorithms to choose from \cite{absalom2022comprehensive}. E.g., if the data distribution is a mixture of Gaussians, the mixture components and mixture weights can be learned with arbitrarily small error, using a number of samples and runtime that are only polynomial in the inverse error \cite{moitra2010settling}. In our implementation, we use the expectation-maximization algorithm readily available in Scikit-learn \cite{scikit-learn}.
In GMMs, the distributions of the \(K\) components and the mixture weights are typically learnt together. To learn the mixture weights in the non\hyp participant distribution \(D\mid Z=0\), one could employ a Bayesian approach and compute
\begin{align*}
    &\Pr(C=k\mid X^0, Z=1)\\
    &= \frac{\Pr(C=k\mid Z=1)\prod_{d\in X^0} \Pr(D=d\mid C=k, Z=1)}
    {\sum_l \Pr(C=l\mid Z=1) \prod_{d\in X^0} \Pr(D=d\mid C=l, Z=1)},
\end{align*}
with the weights \(\Pr(C=k\mid Z=1)\) learnt from the participant data as the prior. However, the records in \(X^0\) are distributed over the different non\hyp participants, and we want to only collect a very small amount of information from each non\hyp participant, which precludes this option. The same holds for estimating the mixture weights using the expectation-maximization algorithm, since this would require multiple rounds of data exchange with the users.
Instead, as described in the following two paragraphs, we let each non\hyp participant send a single class index sampled according to the posterior distribution over classes given this user's data point \(d\), which we use as the vector \(\rho^d\) \cite[Ch. 21.4.1]{murphy2022probabilistic}:
\begin{align*}
    &\Pr(C=k\mid D=d, Z=1)\\
    &= \frac{\Pr(C=k, Z=1)\Pr(D=d\mid C=k, Z=1)}
    {\sum_l \Pr(C=l\mid Z=1) \Pr(D=d\mid C=l, Z=1)}.
\end{align*}
By counting from how many users we receive each index, we can then compute the estimate
\begin{align}
\begin{split}
\label{eq:bayes_approx_expected_posterior}
    &\mathbb{E}_{d\sim D\mid Z=0} \Pr(C=k\mid D=d, Z=1)\\
    &\approx \frac{1}{|X^0|} \sum_{d\in X^0} \Pr(C=k\mid D=d, Z=1)\\
    &\approx \frac{1}{|X^0|} \sum_{d\in X^0} \rho^d_k,
\end{split}
\end{align}
which can be seen as a single Bayesian update step for the distribution of \(C\), using the entire distribution \(D\mid Z=0\). We use this as our estimate for \(\Pr(C=k\mid Z=0)\).
If we wanted to instead perform \(k\leq |X^0|\) Bayesian update steps, we could split the non\hyp participants into \(k\) batches of \(|X^0|/k\) users. We would compute the update in \eqnref{eq:bayes_approx_expected_posterior} for the users in the first batch. Then we would send the resulting posterior for \(C\) to the users in the second batch, and compute the update in \eqnref{eq:bayes_approx_expected_posterior} for these users, but now with the previously computed posterior as the prior for \(C\), use the newly updated distribution as the prior for the users in the third batch, and so on.
This would mean that we could not collect data from all non\hyp participants in parallel, but only in batches. Also, there is a trade\hyp off between the number of update steps and the accuracy in approximating the expectation via \eqnref{eq:bayes_approx_expected_posterior}.

\xhdr{The exponential mechanism}
The exponential mechanism \cite{mcsherry2007mechanism} used in Step~\ref{enum:exponential} is an algorithm whose outputs fulfill (local) differential privacy (see the definition in the introduction). It gets an input dataset \(x\) and is parametrized by a utility function \(u\) that describes the utility of each potential output given an input. It returns a single value \(r\) out of some range \(\mathcal{R}\) with probability proportional to \(\exp(\frac{\varepsilon u(x, r)}{2\Delta u})\), where \(\Delta u\) is the sensitivity of \(u\), that is, how much the value of \(u\) can change at most due to a change in a single record in the input database. In our setting, for a given point \(d\), \(x = \rho^d\) (the dataset consists only of a single record), and \(\mathcal{R} = \{1,\dots,K\}\), i.e., the different possible classes. The utility function is given by \(u(\rho^d, r) = \rho^d_r\), meaning the utility of each class is the probability of \(d\) lying in it. Because \(\rho^d\) is a probability vector and thus \(0\leq u(\rho^d, r) \leq 1\) for any \(\rho^d\) and any \(r\), the sensitivity of \(u\) is 1. We hence sample proportional to \(\exp(\frac{\varepsilon u(x, r)}{2})\), which is a weighted softmax. From the resulting histogram we would like to estimate the distribution of \(C\mid Z=0\).

\xhdr{Postprocessing of the class counts}
If the classes sent by the users were directly sampled from \(\rho^d\), we could simply count how often each of the values \(1,\dots,K\) has been sent to get the estimate of \(C\mid Z=0\) from \eqnref{eq:bayes_approx_expected_posterior}. However, since we perform sampling via the exponential mechanism, we need to revert the distortion introduced by the weighted softmax function. Let
\begin{equation*}
    p=\mathbb{E}_{d\sim D\mid Z=0}(\rho^d)
\end{equation*}
be the probability vector describing the approximate probability distribution of \(C\mid Z=0\) according to \eqnref{eq:bayes_approx_expected_posterior}. Via the exponential mechanism in Step~\ref{enum:exponential} we get \(|X^0|\) i.i.d.\ samples \(A_1,\dots,A_{|X^0|}\) with values in \(1,\dots,K\), and with distribution
\begin{equation*}
    \Pr(A_i = k) = \mathbb{E}_{d\sim D\mid Z=0} \left[\frac{e^{\varepsilon \rho^d_k/2}}{\sum_{l=1}^K e^{\varepsilon \rho^d_l/2}}\right].
\end{equation*}
A first\hyp order Taylor approximation around the mean of \(\rho^d\) \cite[Ch. 4.3.3]{benaroya2005probability} gives us
\begin{align*}
    \Pr(A_i = k) &\approx \frac{e^{\varepsilon \mathbb{E}_{d\sim D\mid Z=0} \rho^d_k/2}}{\sum_{l=1}^K e^{\varepsilon \mathbb{E}_{d\sim D\mid Z=0} \rho^d_l/2}}\\
    &= \frac{e^{\varepsilon p_k/2}}{\sum_{l=1}^K e^{\varepsilon p_l/2}}.
\end{align*}
Let \(\tilde{U}_k = |\{i:A_i=k\}|\), i.e., a random variable counting how often class \(k\) has occurred. Define \(U_k\), for \(k,l=1,\dots,K\), via
\begin{equation}
    \label{eq:Uk}
    U_k - U_l = \frac{2}{\varepsilon}\log\left(\frac{\tilde{U}_k}{\tilde{U}_l}\right)
\end{equation}
and \(\sum_{k=1}^K U_k = 1\). We can compute \(U_1 = \frac{1}{K} (1 + \sum_{k=2}^K (U_1 - U_k))\) and from there \(U_2,\dots,U_K\). \(U_1,\dots,U_K\) are the desired approximations of \(\Pr(C=1\mid Z=0),\dots,\Pr(C=K\mid Z=0)\). We can see this as follows.
Due to the central limit theorem, \(\frac{1}{|X^0|}(\tilde{U}_1,\dots,\tilde{U}_K)\) is an asymptotically normal and consistent estimator of \((\Pr(A_i=1),\dots,\Pr(A_i=K))\). Thus, due to the delta method \cite{cox2005delta}, the expression in \eqnref{eq:Uk} is asymptotically normal as well, and is a consistent estimator of
\begin{align*}
    \frac{2}{\varepsilon}\log\left(\frac{\Pr(A_i=k)}{\Pr(A_i=l)}\right) &\approx \frac{2}{\varepsilon}\log\left(\frac{p_k}{p_l}\right)\\
    &= p_k - p_l.
\end{align*}
Therefore, \(U_k\) is an approximation of \(p_k\), which is an approximation of \(\Pr(C=k\mid Z=0)\).

\xhdr{Approximating the propensity scores}
We next define the \emph{cluster propensity score} \(\tilde{e}\), which can be interpreted as the propensity score of points that lie in exactly one cluster \(k\), i.e., whose cluster probability distribution is binary:
\begin{align*}
    \tilde{e}(k) &= \Pr(Z=1\mid C=k)\\
    &=\frac{\Pr(C=k\mid Z=1)\Pr(Z=1)}{\Pr(C=k)}\\
    &=\frac{\sum_d \Pr(D=d, C=k\mid Z=1)\Pr(Z=1)}{\sum_{d}\Pr(D=d, C=k)}\\
    &\approx\frac{\frac{\sum_{d\in X^1} \rho^d(k)}{|X^1|} \frac{|X^1|}{|X^0| + |X^1|}}{\frac{\sum_{d\in X^1} \rho^d(k) + U_k |X^0|}{|X^0| + |X^1|}}\\
    &= \frac{\sum_{d\in X^1} \rho^d(k)}{\sum_{d\in X^1} \rho^d(k) + U_k |X^0|}.
\end{align*}
With the cluster propensity scores, we are finally able to compute an approximation of the propensity score \(e(d)\) as
\begin{align*}
    e(d) &= \Pr(Z=1\mid D=d)\\
    &=\sum_{k=1}^K \Pr(Z=1\mid C=k,D=d)\Pr(C=k\mid D=d)\\
    &=\sum_{k=1}^K \Pr(Z=1\mid C=k)\Pr(C=k\mid D=d)\\
    &= \sum_{k=1}^K \tilde{e}(k) \rho^d_k.
\end{align*}

\begin{figure*}[t!]
		\centering 
        \includegraphics[width=0.8\textwidth]{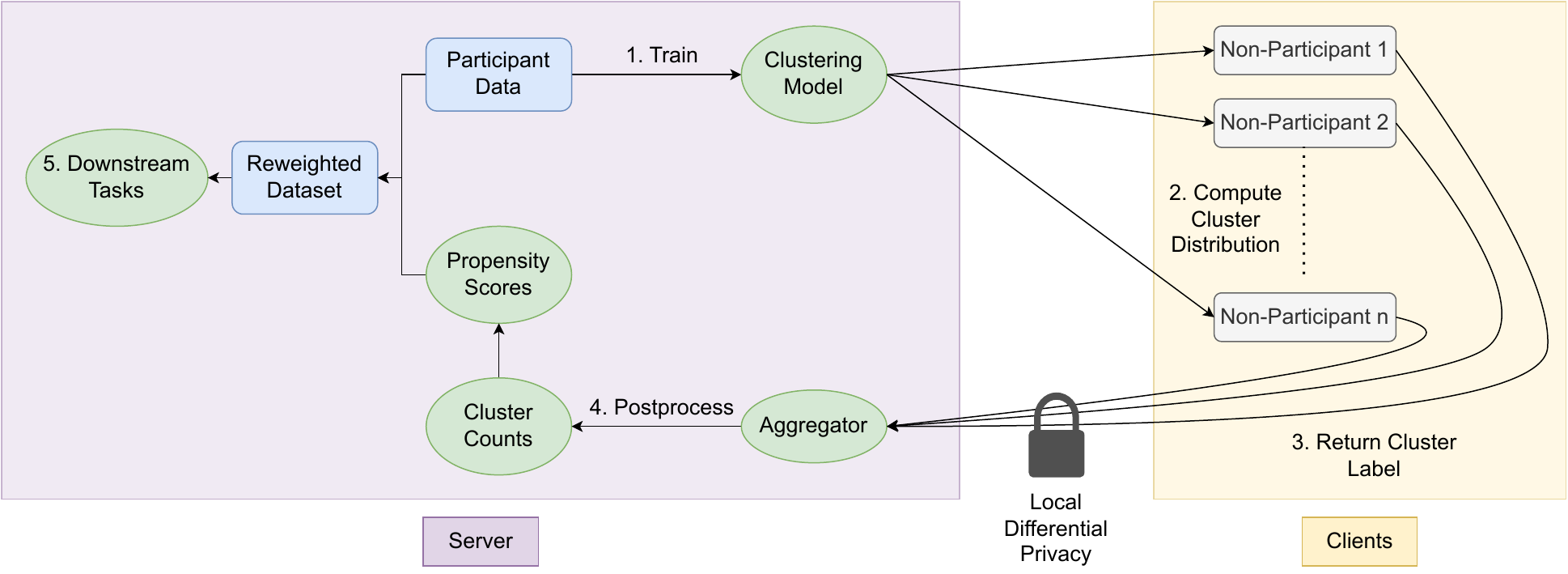} 
  \caption{Overview of \textit{DiPPS}. The party interested in analyzing individuals' data (``Server'') has access to data from individuals that are less concerned about their privacy (``Participant Data'') --- or to proxy data ---, but not to data from individuals that are concerned about their privacy (``Clients''). The server wants to remove the resulting participation bias. To this end, it trains a clustering model on the participant data and ships it to the clients. The clients compute a cluster distribution for their data records using the model, sample one cluster label from that distribution in a locally differentially private way and send it to the server. The server uses the cluster label distribution to compute propensity scores for participation, which are used to reweight the original dataset to remove its bias. This reweighted dataset can then be used for downstream tasks such as statistical analyses.}
        \label{fig:diagram}
\end{figure*}

\Figref{fig:diagram} shows an overview of our method.

\section{Experiments}
\label{sec:experiments}
To evaluate how well \textit{DiPPS} works in practice, we perform several experiments on four very different datasets.\footnote{The datasets are public and the code for reproducing the experiments is available at \url{https://github.com/epfl-dlab/DIPPS}.} We only evaluate the utility of \textit{DiPPS} but not the privacy it provides, since the privacy stems from the exponential mechanism, whose privacy properties have been proven analytically \cite{mcsherry2007mechanism}. Privacy for the non\hyp participants is thus information\hyp theoretically guaranteed, even for worst\hyp case datasets.

\subsection{Tasks}
The goal of \textit{DiPPS} is to approximate a probability distribution. Hence, it is most natural to evaluate it w.r.t.\ a metric for the distance between distributions. As the metric we choose the Wasserstein distance of order 1.
It measures how expensive it is to transform one distribution into the other, when the cost of transporting a given amount of probability mass between two points is the transported amount times the Euclidean distance between the two points. As compared to, e.g., the Kullback\hyp Leibler divergence, the Wasserstein distance does not require the two distributions to have the same support, which is important for us, because we work with finite samples from, in most cases, continuous distributions. We compute the Wasserstein distance using the R transport package \cite{transport2020}. Because the computation
--- which is only required for the evaluation, not for the deployment of our method ---
takes quite long, we only compute the distance for \(\varepsilon=1\).

We also evaluate \textit{DiPPS} on several potential downstream tasks. These tasks are the computation of mean, variance and median of each attribute for the entire dataset and for the non\hyp participant dataset. We measure performance in terms of the absolute value of the deviation from the ground truth, divided by the number of features in the dataset, to make different datasets comparable.

We repeat each experiment five times and report the means of the five repetitions. In addition, for \(\varepsilon=1\) we report the standard deviation of the Wasserstein distance (\Tabref{table:was}) and of the error when estimating mean, variance and median (Tables~\ref{table:mean_err}--\subref{table:median_err}).

\subsection{Datasets}
In this subsection, we evaluate \textit{DiPPS} and its competitors on several datasets. We normalize all features to lie in the interval \([-1,1]\) to make the experiment results comparable. Each dataset is split according to one of its variables into participant and non\hyp participant data.

\xhdr{Web visits} \cite{Dua:2020}
This dataset contains web traces of different users of msnbc.com within a 24h period. The website visits are recorded at the level of 17 URL categories such as ``tech'' or ``business''. We convert this dataset into one that contains for each user whether they have visited a URL of a category zero times, one time or more than once. We cannot use the original unbounded counts, because the \textit{Laplace} and the \textit{Hybrid} mechanism that we compare with require attributes in a bounded range.\\
\textbf{Splitting}: We split the dataset on the ``bulletin board service'' (bbs) attribute. Users who used a bbs at least once are treated as participants, users who never used one are treated as non\hyp participants. The setting could be that of a provider of an Internet platform that is able to record the web visits of their users within the platform but no web visits outside of the platform.
The dataset contains 2k records for which the bbs value is non\hyp zero. From the remaining 988k records, we sample 2k at random as the non\hyp participants.

\xhdr{Credit cards} \cite{yeh2009comparisons}
The dataset contains information about credit card customers. This includes demographic information such as age and gender, and information about their payment history, that is, their repayment status, their previous payments and their bill statements in previous months. In total there are 24 attributes.\\
\textbf{Splitting}: We split the dataset according to the binary variable whether someone defaulted on their credit card payments or not. Participants are those who did not default, non\hyp participants those who did. The reasoning is that people who are struggling financially might be less willing to disclose information about their finances. There are 23.5k participants and 6.5k non\hyp participants.

\xhdr{Food} \cite{foodaps2016}
This data stems from the National Household Food Acquisition and Purchase Survey of the U.S.\ Department of Agriculture, in which data about food purchases of households was collected. We choose a subset of four attributes that are the answers to questions about whether respondents could afford enough food and the food they wanted in the past 30 days.\\
\textbf{Splitting}: For splitting we choose a fifth variable that indicates whether anyone in the households receives benefits from the Supplemental Nutrition Assistance Program, a government program to financially support food purchases. Those who receive benefits are participants, those who do not receive benefits are non\hyp participants. It could, e.g., be that people who are supported by a nutrition assistance program have to provide data about their food purchases to receive the benefits, while others do not. There are 1.5k participants and 3k non\hyp participants in the dataset.

\xhdr{Weather} \cite{zhang2017cautionary}
The dataset contains hourly air pollutant and weather data from twelve air\hyp quality monitoring sites in the Beijing area over a span of four years. The air pollutant attributes include PM2.5 and NO2 concentration, the weather attributes include temperature and precipitation. In total there are twelve attributes.\\
\textbf{Splitting}: We use Wanliu, a site in the city center of Beijing and Dingling, a site in a more rural area nearby. We perform two kinds of experiments for this dataset: Once the city data is treated as the participant data and the rural data as the non\hyp participant data (``Weather 1''), and once it is the other way around (``Weather 1 Inverse''). One could imagine a scenario where an organization has a weather station in one region of a country, but would like to also gather weather information from a different region. The data collected from the weather station would be the participant data. Individuals from the other region of interest could donate weather data that they have collected in a differentially private way; this would then be the non\hyp participant data. To show that our results are robust w.r.t.\ the chosen sites, we repeat the experiments with the sites Nongzhanguan (city center) and Huairou (rural area); we denote those experiments with ``Weather 2'' and ``Weather 2 Inverse''. For each site there are between 30.5k and 33k records after the removal of records with missing data.

\subsection{Comparison Methods}
\label{sec:baselines}
To the best of our knowledge, \textit{DiPPS} is the first method for differentially private distribution estimation in the data setting that we consider. Therefore, there does not exist any method that we can directly compare with (apart from naive estimation, see below). Instead, we compare with a selection of more specialized methods that are limited to computing a small set of functions. As opposed to that, with \textit{DiPPS} any number of arbitrary functions of the data distribution can be computed. This comparison is just to give an idea of the performance of \textit{DiPPS}; we do \emph{not} aim at improving upon the state\hyp of\hyp the\hyp art in those specialized tasks.

\xhdr{Naive}
Compute all functions on the participant data and ignore the non\hyp participant data. This measures the amount of participation bias in the data.

\xhdr{Propensity scores (\textit{PS})}
A strong ceiling: our method, but without the exponential mechanism, i.e., without a differential privacy guarantee.
Instead of sampling via the exponential mechanism, the non\hyp participants directly sample from the cluster distribution.

\xhdr{Laplace mechanism} \cite{dwork2006calibrating}
The \textit{Laplace} mechanism adds noise from a Laplace distribution to each non\hyp participant record and shares these noisy records with the data collector. We merge the resulting noisy dataset with the non\hyp noisy dataset of the participants. The \textit{Laplace} mechanism can be used to estimate mean and median.

\xhdr{Hybrid mechanism} \cite{wang2019collecting}
The \textit{Hybrid} mechanism is a state\hyp of\hyp the\hyp art mechanism for locally differentially private mean estimation. It works similar to the \textit{Laplace} mechanism, but instead of adding unbounded noise to the data, it returns a random value from a bounded interval based on its input. It combines the piecewise mechanism introduced in the same paper with the earlier Duchi mechanism \cite{duchi2018minimax}, and improves upon them in terms of worst\hyp case noise variance. While theoretically it could also be used to estimate the median, the \textit{Hybrid} mechanism is designed for estimating the mean and indeed performs much worse when estimating the median than even the naive method, and hence we only include it for comparison in the estimation of the mean.

\subsection{Hyperparameters of \textit{DiPPS}}
Unfortunately, we cannot choose the best hyperparameters for the downstream task, because in the real\hyp world setting we do not have access to the data of the non\hyp participants and thus no ground truth. We hence resort to choosing the number of PCA dimensions such that at least 80\% of the variance is retained, and the number of GMM components based on the GMM log\hyp likelihood using the elbow method \cite{thorndike1953belongs}. Alterntively, one could choose the number of components based on a regularized log\hyp likelihood that penalizes large numbers of model variables, as in the Bayesian information criterion \cite{schwarz1978estimating} or the Akaike information criterion \cite{akaike1974new}.

\subsection{Results}
\begin{table}
\begin{tabular}{llll}
\hline
                 & \textit{Naive} & \textit{PS}  & \textit{DiPPS}            \\ \hline
Web Visits       & 1.450 & 1.071 & \textbf{1.172 $\pm$ 0.107} \\ 
Credit Cards     &  0.519     & 0.339  &    \textbf{0.467 $\pm$ 0.074}               \\ 
Food             & 0.811 & 0.062 & \textbf{0.531 $\pm$ 0.103} \\ 
Weather 1          & 0.313 & 0.274 & \textbf{0.302 $\pm$ 0.028} \\ 
Weather 1 Inv & 0.313 & 0.285 & \textbf{0.302 $\pm$ 0.013} \\ 
Weather 2          & 0.357 & 0.288 & \textbf{0.291 $\pm$ 0.010} \\ 
Weather 2 Inv & 0.357 & 0.302 & \textbf{0.325 $\pm$ 0.024} \\ \hline
\end{tabular}
\caption{Euclidean norm Wasserstein distance between estimated non\hyp participant data distribution and true non\hyp participant data distribution for \(\varepsilon=1\). Averaged over five runs. Best value in each row is in bold (excluding \textit{PS}).}
\label{table:was} 
\end{table}
\xhdr{Wasserstein distance}
In \Tabref{table:was} we show the difference in Wasserstein distance with Euclidean norm cost between the non\hyp participant distribution estimated by \textit{PS} and \textit{DiPPS} (for \(\varepsilon=1\)) and true non\hyp participant distribution, and compare this with the distance between the participant distribution and the non\hyp participant distribution (\textit{Naive}). Note that we do not show the results for the distance between the estimated and true entire distribution, because they can be computed from the given results by multiplying the distance by \(|X^0|/(|X^0| + |X^1|)\). As a consequence, whenever one method outperforms the other for estimating the non\hyp participant distribution, it will also outperform that method for estimating the entire distribution.
We can see from the table that \textit{DiPPS} outperforms \textit{Naive} on all datasets, which shows that our method indeed successfully reduces participation bias as measured in Wasserstein distance. The Wasserstein distance is widely used, because it captures differences between distributions well. Hence, a reduction in Wasserstein distance as achieved by \textit{DiPPS} will most likely be correlated with a reduction in participation bias in many downstream tasks. We confirm this hypothesis in the next subsection. To show the cost of privacy, we also include the non\hyp private \textit{PS} in the table.

\begin{figure*}[!htbp]
\begin{subfigure}{\textwidth}
		\centering
		\includegraphics[width=0.998\textwidth]{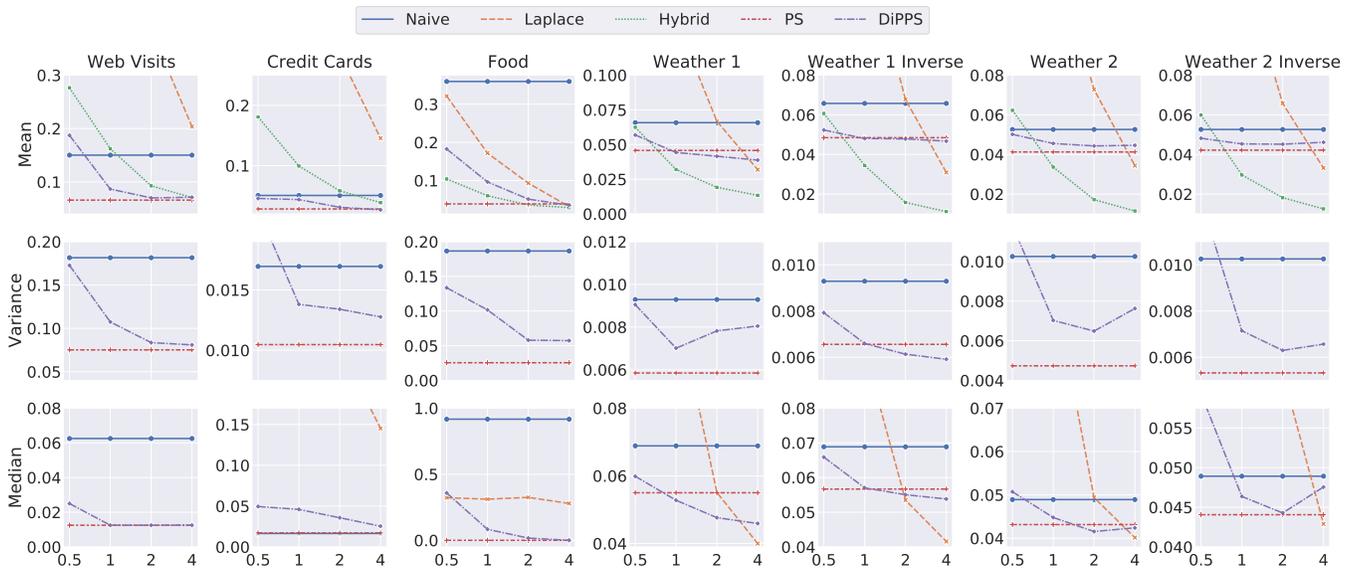}
  \subcaption{Error for non\hyp participant data.}
        \label{fig:stats_nonopt}
\end{subfigure}

\vspace*{1em}

\begin{subfigure}{\textwidth}
		\centering 
        \includegraphics[width=0.998\textwidth]{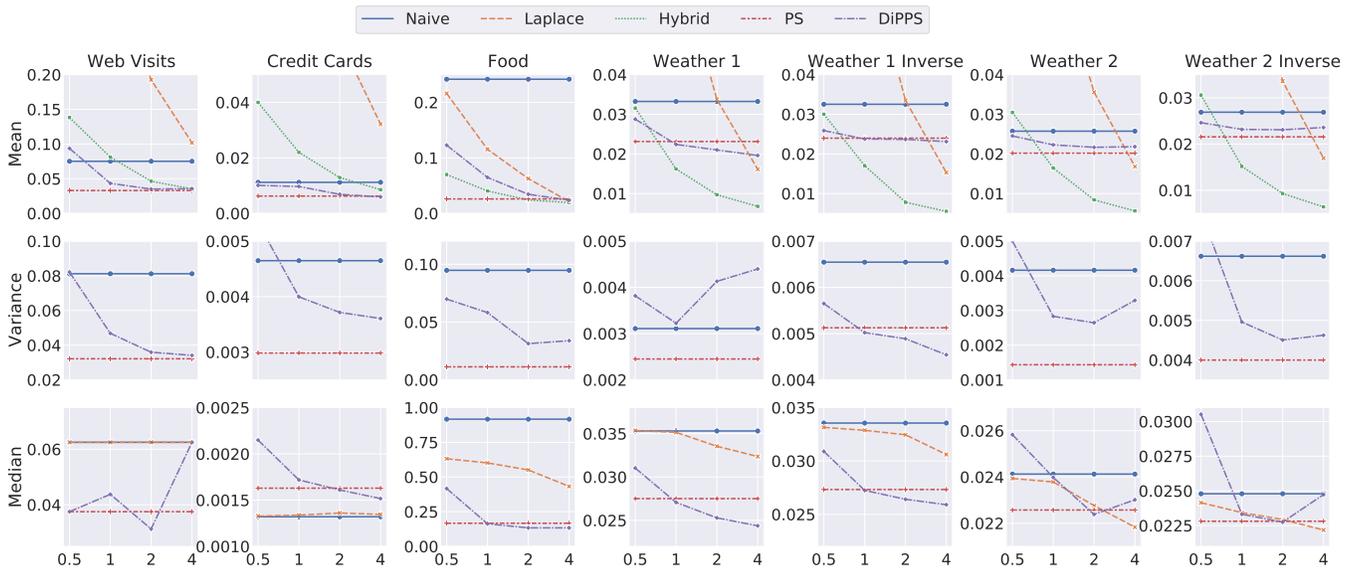}
  \subcaption{Error for entire data.} 
        \label{fig:stats_all}
\end{subfigure}
\caption{\boldmath Mean absolute error (\(=2\ \times\) relative error) per attribute when estimating mean, variance and median of the non\hyp participant (\subref{fig:stats_nonopt}) and entire (\subref{fig:stats_all}) data of different datasets for different $\varepsilon$; lower is better. Averaged over five runs.\unboldmath}
\end{figure*}

\begin{table*}
\begin{subtable}{\textwidth}
\centering

\resizebox{\textwidth}{!}{\begin{tabular}{llllllll}
        & Web Visits                          & Credit Cards                          & Food                          & Weather 1                       & Weather 1 Inv      & Weather 2                       & Weather 2 Inv      \\ \hline
\textit{Naive}   & 0.150                        & 0.051                        & 0.360                         &      0.066                    &   0.066           & 0.053  & 0.053 \\ \hline
\textit{Laplace} & 0.732 $\pm$ 0.124            & 0.603 $\pm$ 0.074            & 0.172 $\pm$ 0.025             &     0.139 $\pm$ 0.028         &    0.146 $\pm$ 0.038  & 0.142 $\pm$ 0.035  & 0.136  $\pm$ 0.040 \\
\textit{Hybrid}  & 0.162 $\pm$ 0.014            & 0.099 $\pm$ 0.014            & \textbf{0.060 $\pm$ 0.010}    & \textbf{0.032 $\pm$ 0.005}    &    \textbf{0.034 $\pm$ 0.003}  & \textbf{0.034 $\pm$ 0.003}  & \textbf{0.030 $\pm$ 0.003} \\
\textit{DiPPS}  & \textbf{0.087 $\pm$ 0.008}             & \textbf{0.044 $\pm$ 0.017}   & 0.096 $\pm$ 0.040             &    0.044 $\pm$ 0.004          &     0.048 $\pm$ 0.001  & 0.045 $\pm$ 0.003  & 0.045 $\pm$ 0.003 \\ \hline
\textit{PS}    & 0.066                        & 0.028                        & 0.039                         &  0.046                        &       0.049          & 0.041  & 0.042
\end{tabular}}
\subcaption{Error when estimating mean.}
\label{table:mean_err}
\end{subtable}

\begin{subtable}{\textwidth}
\centering

\resizebox{\textwidth}{!}{\begin{tabular}{llllllll}
        & Web Visits                        & Credit Cards                          & Food                          & Weather 1                       & Weather 1 Inv      & Weather 2                       & Weather 2 Inv      \\ \hline
\textit{Naive}   & 0.182                       & 0.017                       & 0.187                         &      0.009                    &   0.009           & 0.010  & 0.010 \\ \hline
\textit{DiPPS}  & \textbf{0.107 $\pm$ 0.034}             & \textbf{0.014 $\pm$ 0.002}   & \textbf{0.102 $\pm$ 0.070}             &    \textbf{0.007 $\pm$ 0.001}          &     \textbf{0.007 $\pm$ 0.001}  & \textbf{0.007 $\pm$ 0.002}  & \textbf{0.007 $\pm$ 0.001} \\ \hline
\textit{PS}    & 0.075                        & 0.011                        & 0.027                         &  0.006                        &       0.007          & 0.005  & 0.005
\end{tabular}}
\subcaption{Error when estimating variance.} 
\label{table:var_err}
\end{subtable}

\begin{subtable}{\textwidth}
\centering

\resizebox{\textwidth}{!}{\begin{tabular}{llllllll}
        & Web Visits                          & Credit Cards                          & Food                          & Weather 1                       & Weather 1 Inv      & Weather 2                       & Weather 2 Inv      \\ \hline
\textit{Naive}   & 0.063                       & \textbf{0.016}                       & 0.917                         &      0.069                   &   0.069           & 0.049  & 0.049 \\ \hline
\textit{Laplace} &0.551 $\pm$ 0.069            & 0.450 $\pm$ 0.039            & 0.311 $\pm$ 0.041             &     0.112 $\pm$ 0.014        &    0.097 $\pm$ 0.015  & 0.099 $\pm$ 0.020  & 0.098 $\pm$0.011 \\
\textit{DiPPS}  & \textbf{0.013 $\pm$ 0.028}             & 0.046 $\pm$ 0.043   & \textbf{0.083 $\pm$ 0.059}             &    \textbf{0.053 $\pm$ 0.006}          &     \textbf{0.057 $\pm$ 0.004}  & \textbf{0.045 $\pm$ 0.001}  & \textbf{0.046 $\pm$ 0.004} \\ \hline
\textit{PS}    & 0.013                        & 0.017                        & 0.000                         &  0.055                        &       0.057          & 0.043  & 0.044
\end{tabular}}
\subcaption{Error when estimating median.}
\label{table:median_err}
\end{subtable}

\caption{\boldmath Mean absolute error (\(=2 \times \text{relative error}\)) per attribute when estimating mean (\subref{table:mean_err}), variance (\subref{table:var_err}) and median (\subref{table:median_err}) of the non\hyp participant data with \(\varepsilon=1\). Averaged over five runs. Optimal values in each column are in bold (excluding \textit{PS}).\unboldmath}
\end{table*}

\xhdr{Statistics}
\Figref{fig:stats_nonopt} and \ref{fig:stats_all} show the results for estimating mean, variance and median of the non\hyp participant dataset and the entire dataset, respectively. Tables~\ref{table:mean_err}--\subref{table:median_err} contain the numerical values for estimating the non\hyp participant statistics at \(\varepsilon=1\). \textit{DiPPS} outperforms the naive estimation for \(\varepsilon \geq 1\) in almost all settings. This confirms our hypothesis that a smaller Wasserstein distance corresponds to a smaller error on downstream tasks. In most settings, \textit{DiPPS} outperforms the \textit{Laplace} mechanism. And even the \textit{Hybrid} mechanism is beaten by \textit{DiPPS} on two out of the five datasets.

We want to emphasize that our main goal is not to improve upon existing locally differentially private mechanisms that are specialized for certain tasks. While those are only suitable for computing one specific function, when using our method \emph{any} function of the dataset can be computed --- without decreasing the privacy guarantee when increasing the number of function computations ---, because \textit{DiPPS} estimates a data distribution instead of a specific function. Hence, \textit{DiPPS} fulfills a much more holistic purpose than specialized mechanisms.

\xhdr{Influence of \boldmath\(\varepsilon\)\unboldmath}
In all plots it can be seen that the errors of the \textit{Laplace} and the \textit{Hybrid} mechanism decrease with increasing \(\varepsilon\). This is expected because the added noise gets smaller the larger \(\varepsilon\) is. However, \textit{DiPPS} does not necessarily behave the same; see, e.g., the variance estimation for the Weather 1 dataset. The reason is that the noise due to the exponential mechanism is more subtle. \(\varepsilon\) influences the sampling process, and the optimal \(\varepsilon\) depends on the data. For example, a very large \(\varepsilon\) would mean that in all but very few cases the class that has the highest probability for a point would get sampled, ignoring the probabilities of the other classes. If all data points have probability close to \(1\) for a single class, this would be desirable. However, if there is, e.g., one class that has probability considerably larger than \(0\) for many points, but for none of the points is the class with the highest probability, this class might not get sampled at all and thus its frequency would get underestimated.
The subtle influence of the exponential mechanism is also the reason why in some cases \textit{DiPPS} outperforms \textit{PS}. The distribution from which our datasets were sampled is most likely not an exact Gaussian mixture model. This means that the approximations of the propensity scores are not perfect. In some cases, the distortions introduced by the exponential mechanism correct part of this error and lead to slightly better results than when sampling directly from the cluster distribution.

\section{Discussion}
\label{sec:discussion}
\xhdr{Practicality of \textit{DiPPS}}
As shown in our experiments, \textit{DiPPS} offers an improvement over a naive estimation of statistics in the large majority of cases, even though this improvement often is only moderate. Hence, when one expects a difference between the distribution of the participant data and the non\hyp participant data, then often the question will not be whether using \textit{DiPPS} will lead to better results. It will rather be whether the additional effort of implementing our method and its overhead will be worth it, which very much depends on the particular use case. Points in favor of implementing \textit{DiPPS} are that it only requires a single round of communication and its holistic nature, meaning that it is not specialized for any particular function computation, but allows for computing arbitrary functions on the data.

\xhdr{Implementation options}
\textit{DiPPS} is a very modular method.
For instance, using a GMM combined with a PCA is only one out of many possible ways how the clustering could be implemented. Other clustering methods could be used, or even general dimensionality reduction methods such as an autoencoder: the latent vector for a data point could be normalized and interpreted as a probability distribution over the different dimensions. For text data, a topic model such as LDA could be used. The clustering component could even be replaced entirely: The propensity scores could be directly computed by an ML model such as logistic regression that is trained via differentially private federated learning \cite{truex2019hybrid}.
For preserving the privacy of the non\hyp participants, we sample a value from their class distributions via the exponential mechanism. It would be interesting to investigate replacing the exponential mechanism with the more recent permute\hyp and\hyp flip mechanism \cite{mckenna2020permute}, which can produce outputs with higher utility function values.
If the class assignment is a hard instead of a soft one, i.e., the clustering model does not return a probability distribution over classes but only a single class, then the problem of estimating \(C\mid Z=0\) becomes a frequency estimation problem. So instead of the exponential mechanism, one could use a mechanism for locally differentially private frequency estimation \cite{murakami2019utility}.

\xhdr{Privacy of participants}
In our experiments, the PCA and the GMM are trained in a non\hyp private way on the participant data, and hence do not provide differential privacy for the participant data. This is not an issue for proxy datasets from public sources. For user data it becomes an issue if participants are willing to share their data with the data collecting party but not with other parties. In these cases, one can train a differentially private clustering model instead. There are, e.g., mechanisms for differentially private PCA \cite{wishart2016,chaudhuri2013near}, GMM \cite{kamath2019differentially,park2017dp} or $k$\hyp means \cite{stemmer2018differentially,su2016differentially}.

\xhdr{Broader impact and ethical considerations}
\textit{DiPPS} allows data analysts to perform analyses on data that was previously inaccessible due to privacy concerns. This can help get insights into understudied domains. Even more importantly, \textit{DiPPS} can prevent drawing wrong conclusions from datasets that suffer from participation bias.
However, the privacy guarantees of \textit{DiPPS} are based on differential privacy, for which the privacy parameter \(\varepsilon\) needs to be set. In order to fully understand the privacy implications of their data sharing, individuals must be educated on the meaning of \(\varepsilon\). A malicious data collector might even promise to use a certain value for \(\varepsilon\), but in the implementation choose a larger \(\varepsilon\); or just send data without any privacy protection at all. This can be prevented by releasing the client code. 
The privacy properties only depend on the client code but are independent of the server, and hence the server code need not be accessible.

\section{Conclusion}
\label{sec:conclusion}
In this paper we have presented Differentially Private Propensity Scores for Bias Correction (\textit{DiPPS}), a method for reducing participation bias or bias resulting from using proxy datasets. It can be used whenever individuals who are not willing to share their raw data are willing to at least share a single locally differentially private value. As opposed to other locally differentially private methods, \textit{DiPPS} estimates a distribution on which arbitrary functions instead of only single specialized ones can be computed. In experiments on datasets from very different domains we have shown that \textit{DiPPS} can indeed reduce bias, and in the estimation of statistics can even outperform locally differentially private methods that are specialized for the given task. We have further pointed out multiple ways in which \textit{DiPPS} can be extended to yet more settings than those covered in this paper.

\bibliography{references}

\end{document}